\documentclass[aps,reprint,twocolumn,superscriptaddress,citeautoscript,showpacs,amsmath,amssymb,floatfix,prl]{revtex4-1}

\usepackage{graphicx}
\usepackage[usenames,dvipsnames]{xcolor}
\usepackage{epstopdf}
\usepackage{amsmath,amsthm,amssymb}
\usepackage{latexsym}
\usepackage{bm}
\usepackage{dcolumn}
\usepackage{braket}
\usepackage[normalem]{ulem}
\usepackage{times}
\usepackage{hyperref}

\begin{document}
\bibliographystyle{apsrev4-1}

\title{Comment on ``Colossal Pressure-Induced Softening in Scandium Fluoride''}
\author{I. A. Zaliznyak}
\email{zaliznyak@bnl.gov}
\author{E. Bozin}
%\email{bozin@bnl.gov}
\affiliation {Condensed Matter Physics and Materials Science Division, Brookhaven National Laboratory, Upton, NY 11973, USA}
\author{A. V. Tkachenko}
%\email{oleksiyt@bnl.gov}
\affiliation{CFN, Brookhaven National Laboratory, Upton, New York 11973, USA}

\date{\today}

%\begin{abstract}
%No abstract needed for a Comment.
%\end{abstract}

\maketitle

In a recent Letter \cite{Wei_etal_Dove_PRL2020}, Wei \emph{et~al.} report neutron powder diffraction measurements at variable temperature and pressure of negative thermal expansion material, scandium fluoride ScF$_3$ \cite{Greve_JACS2010}. The diffraction patterns were fitted using the Rietveld method to refine the lattice and the atomic displacement parameters, with other aspects of the crystal structure fixed by $Pm\bar{3}m$ cubic symmetry. From the structural refinement of the measured diffraction patterns, authors obtain the isothermal compressibility curves. These results thoroughly characterize the equation of state of this material in the low-pressure cubic phase (with increasing pressure, ScF$_3$ undergoes a structural transition to the rhombohedral structure).
%By fitting the results to a phenomenological equation of state, authors observe marked (colossal) pressure-induced softening in this material. , $V = V(P,T)$,

The results reported by Wei \emph{et~al.} \cite{Wei_etal_Dove_PRL2020} are indeed very interesting and important
% and deservedly published in Physical Review Letters
because these results can be confronted with predictive, quantitative theories of NTE and pressure-induced softening and allow to corroborate, or invalidate certain approaches. Wei \emph{et~al.} discuss their observations in the context of model molecular dynamics simulations and simple one-dimensional models, which capture qualitative features of the observed phenomena, but do not provide a quantitative theory. On the other hand, a microscopic theory of vibrational and thermomechanical properties of empty perovskite crystals with ReO$_3$ structure (also rooted in neutron diffraction results \cite{Wendt_2019}) has recently been proposed. This theory describes empty perovslkite structures with strong nearest-neighbor bonds as Coulomb Floppy Networks (CFNs, floppy networks of rigid links stabilized by Coulomb interaction) and provides a very accurate quantitative description of NTE in ScF$_3$ \cite{Wendt_2019,TkachenkoZaliznyak_arXiv2019}.

\begin{figure}[t!]
\includegraphics[width=0.9\columnwidth]{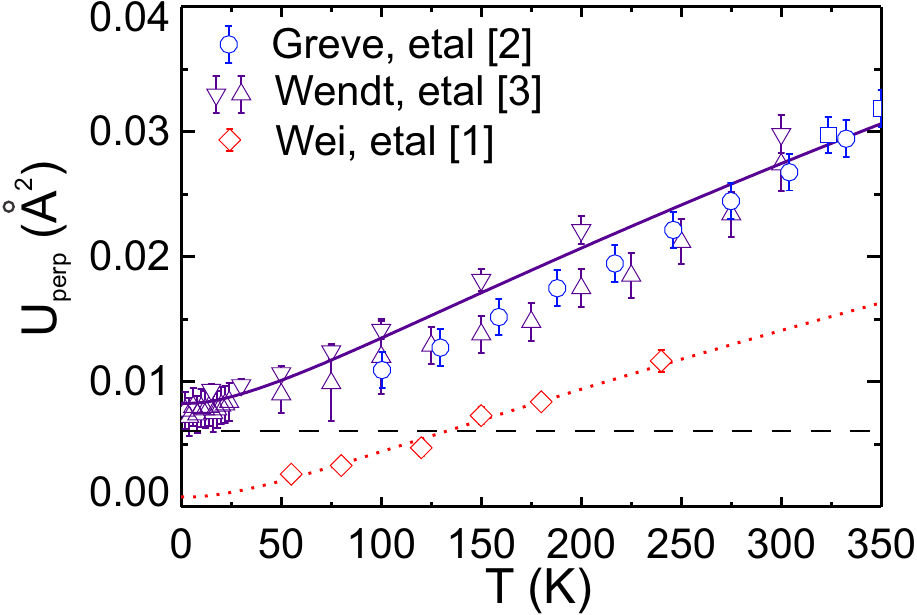}
\caption{Mean-squared transverse displacement of the F atoms obtained from refinement of the crystal structure at P~$\approx0$. The triangles show the results of Rietveld refinement of neutron diffraction data measured on NPDF diffractometer at LANSCE (down triangles) and NOMAD diffractometer at Spallation Neutron Source (up triangles) from Fig. 4b of Wendt, \emph{et~al.} \cite{Wendt_2019}. The circles and squares correspond to the Xray data from Supplementary Figure 1 of Greve, \emph{et~al.} \cite{Greve_JACS2010}. The (red) diamonds are the data of Wei \emph{et~al.} \cite{Wei_etal_Dove_PRL2020}. The solid line is theoretical prediction of Ref.~\onlinecite{TkachenkoZaliznyak_arXiv2019}; the horizontal broken line shows the quantum limit for the Fluorine mean-squared transverse displacement due to zero-point motion at T = 0.}
\label{Fig1}
\vspace{-2mm}
\end{figure}

Motivated to further corroborate the CFN theory, we compared its prediction for the mean-squared transverse displacement of the F atoms, U$_{perp} \equiv \langle u_{perp}^2\rangle$, with that obtained by Wei \emph{et~al.}, and observed a marked discrepancy. The experimental values are much smaller than those expected from theory (Fig.~\ref{Fig1}). In fact, the U$_{perp}$ values of Wei \emph{et~al.} appear unphysically small, falling for T~$\lesssim 150$~K well below the quantum limit for the Fluorine mean-squared transverse displacement due to zero-point motion at T = 0, $\langle u_{0}^2 \rangle = \hbar/m_F\omega_+ \approx 0.006$~\AA$^2$ (dashed line in Fig.~\ref{Fig1}; $\hbar$ is Planck constant, $m_F$ is mass of the F ion, and $\hbar\omega_+ \approx 37$~meV is the transverse F phonon bandwidth). We then compared these results with the previously published X-ray diffraction data of Greve, et al. \cite{Greve_JACS2010} and the neutron diffraction data of Wendt, et al. \cite{Wendt_2019}. We found the latter two data sets to be in a good agreement with each other, as well as with the prediction of CFN theory (Fig.~\ref{Fig1}).

We thus conclude that U$_{perp}$ values reported in Fig.~5 of Ref.~\onlinecite{Wei_etal_Dove_PRL2020} are substantially incorrect. In our experience, such an underestimate of atomic displacement parameters can be caused by an incorrect accounting for the effects of beam absorption in/transmission through the sample and sample environment (such as pressure cell in measurements of Wei \emph{et~al.}) in the Rietveld treatment of the diffraction data. This would also explain unphysical negative atomic displacement parameters reported in Supplementary Figs.~S7 and S8 of Ref.~\onlinecite{Wei_etal_Dove_PRL2020}, where authors indeed write, ``the negative values are consistent with not including the effects of beam attenuation in the refinement process, which at this point could be taken into account by a positive constant shift of all values''. Whether the systematic error of U$_{perp}$ in Fig.~5 of Ref.~\onlinecite{Wei_etal_Dove_PRL2020} can be somehow accounted by a simple shift, is unclear. We believe that the data are of sufficient interest and importance to deserve an analysis better accounting for the absorption/transmission effects, which would eliminate, or markedly reduce the systematic error that is currently present in Fig.~5 of Ref.~\onlinecite{Wei_etal_Dove_PRL2020}.

The purpose of this Comment is twofold: (i) to caution the researchers against using the U$_{perp}$ data of Wei \emph{et~al.} \cite{Wei_etal_Dove_PRL2020} for quantitative comparisons with theory, and (ii) to encourage the authors of Ref.~\onlinecite{Wei_etal_Dove_PRL2020} to reconsider their analysis and obtain a reliable U$_{perp}$ data by better accounting for the beam transmission and attenuation effects.

\begin{acknowledgments}
Work at Brookhaven National Laboratory was supported by Office of Basic Energy Sciences (BES), Division of Materials Sciences and Engineering, U.S. Department of Energy (DOE), under contract DE-SC0012704. Work at BNL's Center for Functional Nanomaterials (CFN) was sponsored by the Scientific User Facilities Division, Office of Basic Energy Sciences, U.S. Department of Energy, under the same contract.
\end{acknowledgments}

%\bibliography{../BibTex_NTE_full}

%\end{document}

%

\end{document}